\documentclass[a4paper]{jpconf}
\usepackage{graphicx}
\begin{document}
\title{Chiral Transport Phenomena and Compact Stars \jpcs}

\author{Cristina Manuel}

\address{ Instituto de Ciencias del Espacio (ICE, CSIC),
C. Can Magrans s.n., 08193 Cerdanyola del Vall\`es, Catalonia, Spain, and Institut d’Estudis Espacials de Catalunya (IEEC), C .Gran Capit\`a 2-4, Ed. Nexus, 08034 Barcelona, Spain }

\ead{cristina.manuel@csic.es}

\begin{abstract}
 I will review the main chiral transport phemomena arising in systems made up of (almost) massless fermions associated to the quantum chiral anomaly. These  quantum effects might have relevant implications in
 compact stars, and  I will review some relevant works that reveal so.
I will also show how a conservation law that has the same form of  the chiral anomaly also emerge in perfect classical fluids, which expresses a conservation law of magnetic, fluid and mixed helicities for isentropic fluids,
and why this should also be relevant in compact stars.
\end{abstract}

\section{Introduction}
\vskip 0.5 cm
Understanding the role of symmetries, and also the violation of a symmetry, has played a pivotal role in  different branches of
physics. More particularly, in the development and construction of the Standard Model of particle physics  the study of chiral symmetry was crucial.
The weak force was found to be only interacting with left-handed fermions (or right-handed antifermions), revealing a breaking of parity.
Electromagnetic interactions were known to be respectful with  the chirality of the charged particles.  However,  at quantum level the chiral symmetry
was found to be  not longer preserved. This discovery had a deep impact on the construction of theoretical particle physics models, as one had to understand the
criteria under which some symmetries could be broken by quantum effects.

In particle physics the
effects of the quantum chiral anomaly come mainly in the explanation of different anomalous decays, such as that
of the neutral pion into two photons. In many body systems, the quantum chiral anomaly has also several relevant macroscopic effects, as it has been found out that it leads to a wide variety of dissipationless transport phenomena that have a relevant impact in the dynamical evolution of chiral systems. In this talk
   I will discuss and review the most relevant aspects of  chiral transport phenomena, and then
focus on why this is also relevant in the study of compact stars. My intention is not to give a
complete review on this topic (check the existing excellent reviews in the literature, see Refs.~\cite{Kharzeev:2013ffa,Kharzeev:2015znc,Landsteiner:2016led,Hidaka:2022dmn,Kamada:2022nyt}), so I apologize if I cannot cover all existing works on this growing topic in this talk. I will rather  pinpoint the most relevant
concepts and focus on some works that reveal its relevance in compact stars.

\section{Chiral magnetic and vortical effects}
\vskip 0.5 cm

Chiral transport phenomena refers to  quantum transport effects associated to the chirality of fermions, and 
related to the quantum chiral anomaly.  One considers the situation of massless, or quasi-massless fermions, as chirality is only a well-defined quantum number in that case, as otherwise, the mass  mixes up the dynamical evolution of the different fermion chiralities.

I will concentrate on discussing the chiral magnetic and vortical effects.
The chiral magnetic effect (CME) \cite{Fukushima:2008xe} is a phenomenon in which a magnetic field generates an
electric current in a conducting material, such as a plasma or a fluid, that contains a population
of fermions with a chiral imbalance.
One intuitive way of understanding the CME comes by realizing that for massless fermions the spin and momentum are either parallel or antiparallel, depending whether those are right-handed or left-handed (the opposite for the antifermions).  In a magnetic field all the spins are aligned, and this implies that  all  particles then move in the direction of the magnetic field. In the presence of a misbalance among chiralities, these  parallel/antiparallel currents  to the magnetic field are not counterbalanced, and this is ultimately what creates the effect. A similar situation occurs in the presence of vorticity, as also spins align with the vorticity,  then there is a current
parallel to the vorticity vector, and  one talks about the chiral vortical effect (CVE).

The CME or CVE currents depend on the chiral chemical potential, the parameter that gives account of the chiral misbalance. That the effects are related to the chiral anomaly can be shown in different ways, but I will highlight 
 that first used in \cite{Sadofyev:2010is}, based on using some effective field theory methods, and valid at zero temperature.
In the Hamiltonian of a system a chemical potential $\mu$  and chiral chemical potential $\mu_5$ enter as  new pieces in the Hamiltonian that go as $ \delta H =  \mu n + \mu_5 n_5$, where  $n/n_5$ are the charge/chiral charges densities. Very often one says that the chemical potentials act as a ficticious zero component of a vector gauge potential.
 In  a moving system with velocity $u^\mu$ those terms would rather be  $ \delta H =  \mu u_\mu j^\mu + \mu_5 u_\mu j^\mu_5$, where $j^\mu, j^\mu_5$ are the 
 (classically conserved) vector and axial vector currents, respectively. Thus, one sees that $\mu u^\mu$ couples to matter as a real vector gauge field potential $A^\mu$ does, while $\mu_5 u^\mu$
 couples as it were an axial vector gauge potential.  We can push these analogies, and then compute the quantum anomaly in the presence of the ficticious vector and axial vector fields. 
At a quantum level, we can then use 
 the (covariant)  quantum anomalies  \cite{Landsteiner:2016led} for the vector and axial currents, 
 assuming that the vector gauge field is   $e A^\mu +  \mu u^\mu$, while the axial vector field is $\mu_5 u^\mu$ 
\begin{eqnarray}
\partial_\mu j^\mu_5 &=& - \frac {1}{4 \pi^2} \epsilon_{\mu \nu \alpha \beta}  \left( \partial^\mu (e  A^\nu + \mu u^\nu) \partial^\alpha (e A^\beta +  \mu u^\beta) + \partial^\mu (\mu_5 u^\nu) \partial^\alpha (\mu_5 u^\beta) \right) \ ,  \\
\partial_\mu j^\mu &=& -\frac {1}{2 \pi^2} \epsilon_{\mu \nu \alpha \beta} \left( \partial^\mu (e  A^\nu + \mu u^\nu) \partial^\alpha ( \mu_5 u^\beta) \right) \ .
\end{eqnarray}
It turns out that all the pieces with a chemical and chiral chemical potential in the quantum anomaly equations might be re-written down by modifying the expressions of the vector and axial-vector currents.
After integrating by parts, and discarding surfaces terms, one can re-write the above equations as 
\begin{eqnarray}
\label{chiral-anomaly}
\partial_\mu \left( j^\mu_5   + \frac{e \mu}{2 \pi^2} B^\mu + \frac{\mu^2+ \mu_5^2}{2 \pi^2} \omega^\mu \right) &=& - \frac {e^2}{16 \pi^2} \epsilon_{\mu \nu \alpha \beta}  F^{\mu \nu} F^{\alpha \beta} \ , \\
\partial_\mu \left( j^\mu    + \frac{e \mu_5}{\pi^2} B^\mu + \frac{\mu \mu_5}{ \pi^2} \omega^\mu \right) &=& 0  \ ,
\label{conserv-totalV}
\end{eqnarray}
where we have defined
\begin{equation} 
 \omega^\mu = \frac 12  \epsilon^{\mu \nu \alpha \beta} u_\nu \partial_\alpha u_\beta \ , \qquad
B^\mu = \frac 12 \epsilon^{\mu \nu \alpha \beta} u_\nu F_{\alpha \beta} \ ,
\end{equation}
 which in the rest frame 
of the system represent the vorticity and magnetic field vectors, respectively. In a moving frame however, the zero component of these vectors
are associacted to the fluid and mixed helicities (we will later on discuss on this).
 We then see that at $T=0$, one could read the conservation law of the electromagnetic current by multiplying by $e$ the r.h.s. of 
Eq.~(\ref{conserv-totalV}), and  thus identify the CME and CVE currents in this way. Please note that induced chiral currents proportional to 
$B^\mu$ and $\omega^\mu$ are also produced, these are called the chiral separation and chiral vortical effects, respectively.

An interesting remark is that all the chiral transport phenomena that originate in the chiral quantum anomaly are dissipationless, and thus they do not imply an increase of entropy.
Also that the effects are not corrected perturbatively, as they rely on the quantum chiral anomaly.

There are several different systems where all these ideas might be applied. In some materials, like the Weyl of Dirac semi-metals, there are quasiparticles that behave as massless fermions. In these systems the
CME has been detected \cite{Li:2014bha}.
In systems at extreme conditions of temperature and/or density, one might expect that  most fermions can be considered as massless
 whenever their mass $m$ is such that $ m \ll T$ and/or $ m \ll \mu$, where $T$ is the temperature. Thus, one can also expect to find chiral transport phenomena in the quark-gluon plasma
phase studied with heavy-ion collisions. Big efforts by the different experimental collaborations of both the LHC and  RHIC have been carried out (see for example \cite{ALICE:2020siw,STAR:2021mii}), but so far it has not been detected, while there are debates on what  the criteria to claim detection should be    \cite{Kharzeev:2022hqz}.
 We can also expect that
these ideas might be relevant in several cosmological and astrophysical scenarios, as we will encounter several extreme conditions where one can take the
ultrarelativistic limit to describe the corresponding quasiparticles.

\section{ Chiral hydrodynamics and chiral kinetic theory}
\vskip 0.5 cm

 Relativistic hydrodynamics has been naturally applied to a variety of cosmological, astrophysical and nuclear physics
scenarios. The hydrodynamical equations are the expressions of the conservations laws of a system. In the
presence of chiral fermions, it thus seems natural to incorporate the quantum chiral anomaly equation in the hydrodynamics \cite{Son:2009tf},
the resulting framework is known as chiral or anomalous hydrodynamics. 

An interesting result is that even if the quantum chiral anomaly requires  the computation of one-loop Feynman diagrams, the famous triangle diagrams, for many-body systems it is possible to give account of it with
semi-classical methods. Chiral or anomalous hydrodynamics might be decuded from the so called chiral kinetic or transport theory (CKT) \cite{Son:2012zy,Stephanov:2012ki,Chen:2012ca}. There are several derivations of CKT, I will focus on those that I am more familiar with.
The physics associated to these chiral imbalanced systems, governed by the chiral quantum anomaly, might be
deduced by incorporating the first quantum corrections to classical transport equations. One can do that by taking the Dirac Hamiltonian, in the presence of electromagnetic fields,
and diagonalize it for the particle and antiparticle degrees of freedom in an expansion in $\hbar$, the Planck constant \cite{Manuel:2014dza}, treating the resulting expression semi-classically. It is possible also to derive an effective field theory that does the same at a quantum field theory level \cite{Carignano:2018gqt}.
Alternatively, one can derive transport equations from quantum field theory, and expand the resulting equations
in $\hbar$ \cite{Hidaka:2016yjf}. There are some subtleties when one uses one framework over the other to derive CKT, which has to do with the semi-classical definition of quasiparticle (and getting rid of 
the so called  {\it Zitterbewegung} effect  \cite{Carignano:2019zsh}),  but I will not enter in discussing this issue here.

The transport equation obeyed by the distribution function $f_p$ associated to a fermion of chirality $\chi$ can be written down as
(see Refs.  \cite{Son:2012zy,Manuel:2014dza} )
\begin{eqnarray}
 \label{chiraltreq}
  \frac{\partial f_p}{\partial t} &+& \left( 1+ e {\bf B} \cdot {\bf \Omega}\right)^{-1} \left\{ \left[  \tilde {\bf v} + e  \ \tilde{\bf E}  \times {\bf \Omega} 
 + e \ {\bf B} (\tilde{\bf v} \cdot {\bf \Omega}) \right]  \cdot \frac{\partial f_p}{\partial {\bf r}}  \right.
 \nonumber  \\
 &+ &  \left. e \left[  \tilde{\bf E}  + \tilde {\bf v}\times {\bf B} + e   {\bf \Omega} \
  ( \tilde{\bf E}  \cdot {\bf B}) \right] \cdot \frac{\partial f_p}{\partial {\bf p}} \right\} =0 \ .
  \end{eqnarray}
Here ${\bf \Omega} = \chi  {\bf p}/p^3$ is the so called Berry-curvature, and we have defined
 \begin{equation}
 \tilde{\bf E}  =  {\bf E}- \frac 1e \frac{\partial \epsilon_{\bf p}}{\partial \bf r}  \ ,
 \qquad
\tilde {\bf v}  =  \frac{\partial \epsilon_{\bf p}}{\partial {\bf p}} \ , 
\end{equation}
where $\epsilon_{\bf p}$ is the particle's energy.
Although we work in natural units, restoring dimensions, it is possible
to check that all pieces that contain the Berry curvature are proportional to $\hbar$, and are pure quantum effecs that correct the classical
terms of the transport equation. The particle density associated to these chiral fermions  reads
  \begin{equation} 
  n = \int \frac{d^3p}{(2\pi )^3} (1 + e\,  {\bf B} \cdot {\bf \Omega}) f_p \ ,
   \end{equation}
while the current reads
\begin{equation}
 {\bf j} = - \int \frac{d^3 p}{(2\pi )^3} \left[ \epsilon_p \frac{\partial f_p}{\partial {\bf p}}  + e  {\bf \Omega} \cdot \frac{\partial f_p}{\partial {\bf p}} \epsilon_p {\bf B} +  
 \epsilon_p  {\bf \Omega} \times \frac{\partial f_p}{\partial {\bf r}}  - e f_p {\bf E} \times {\bf \Omega} \right] \ . 
 \end{equation}

After integrating the kinetic equation  one then obtains
\begin{equation}
\label{pre-anomaly}
 \frac{\partial n}{\partial t} + \nabla \cdot {\bf j} = - e^2  \int \frac{d^3p}{(2\pi)^3} \left( {\bf \Omega} \cdot \frac{\partial f_p}{\partial {\bf p}} \right) {\bf E} \cdot {\bf B} \ .
  \end{equation}
Now considering the two possible chiralities, one can construct both the vector current and chiral current, combing the contributions of the two chiralities.
  Taking an equilibrium form of the distribution function, 
taking into account both particle and antiparticle degrees of freedom, one can then reproduce the quantum chiral anomaly,  Eq.~(\ref{pre-anomaly}), while one can obtain the conservation of the vectorial current. 
 Written in this form, one sees the clear quantum origin of the non-conservation of the current, as  if the quantum corrections to the classical transport equation are neglected, the chiral current  would be  conserved.

It is interesting to study the dynamical evolution of the CME, allowing the electromagnetic fields to be
dynamical. Let us assume that the system is at rest. After integrating the chiral anomaly equation over space in a closed volume $V$, it leads to a quantum conservation law  that relates the chiral fermion density, $Q_5 = \frac 1V \int d^3 x \,n_5(x)$ with the magnetic helicity of the system
\begin{equation}
\frac{d Q_5}{dt} = \frac{  e^2}{2 \pi^2} \frac 1 V \int d^3 x \, {\bf E} \cdot {\bf B} = - \frac{e^2}{2 \pi^2} \frac{d {\cal H}}{dt}   ,
\end{equation}
where
\begin{equation}
 {\cal H} = \frac {1}{V} \int d^3 x \, {\bf A} \cdot {\bf B}  
 \end{equation}
 is the magnetic field density. This quantity is gauge invariant provided that the magnetic field vanishes at, or it is parallel to the boundary of $V$.
The magnetic helicity gives a measure of the linkage and twists of the magnetic field lines. The above equation tells us  that  the chirality of the fermions can be transformed into magnetic helicity, and/or viceversa, and thus generate/destroy different non-trivial topological field configurations.

As the chiral symmetry is only an approximated symmetry, one should also add in the chiral anomaly equation a chirality flipping rate $\Gamma_f$, typically giving account of scattering processes that allow for a change of chirality
due to the existence of a small mass. Thus,  one should rather write

\begin{equation}
\frac{d Q_5}{dt}  + \frac{e^2}{2 \pi^2} \frac{d {\cal H}}{dt}  = - \Gamma_f n_5 \ .
\end{equation}
The chiral anomaly coupled to Maxwell's equations govern the dynamical evolution of the chiral medium. One can view that  the chirality of the fermions can be transformed into magnetic helicity, and/or viceversa, on time scales shorter than $t \ll 1/\Gamma_f$.
An interesting property is that some electromagnetic modes are unstable, and can grow exponentially. Let us explain why. Assuming the presence of a CME current along the $z$ direction of an applied magnetic field, a chiral magnetic instability arises \cite{Joyce:1997uy,Akamatsu:2013pjd}.  According to Amp\`ere's law, a magnetic field is generated in the $\theta$ direction. This induced field, in turn, generates a current in the same direction through the CME effect, which would
then generate a field in the $z$ direction. This process results in an amplified value of the initial magnetic field in the $z$ direction, leading to a runaway mechanism that causes the instability. The unstable modes occur for field wavelengths which are less than $k \sim e^2 \mu_5$. The time scale of growth
in an electromagnetic plasma is of the order $ \sim 1/e^4 \mu_5$ \cite{Akamatsu:2013pjd}, while for a conductor, with electrical conductivity $\sigma$ it is 
$t_{\rm ins} \sim \sigma/k e^2 \mu_5$ \cite{Boyarsky:2011uy}. In the presence of an initial chiral chemical potential there can be a generation of magnetic helicity
with a sort of inverse cascade phenomenon, when  the helicity is transferred from the highest to the lowest modes \cite{Boyarsky:2011uy}.

In the presence of a small fermion mass, the chiral anomaly equation is also affected by the presence of quantum coherent mixtures of mixed helicities, as seen in Ref.\cite{Manuel:2021oah}. The effect of these genuine quantum states has, however, not yet been studied.

\section{Chiral anomaly equation and compact stars}
\vskip 0.5 cm

A couple of weeks before giving this talk, a review article on chiral transport phenomena in astrophysics and  cosmology appeared in the arxives
 \cite{Kamada:2022nyt},  which contains an exhaustive lists of references on this topic. I cannot cover all works in this talk, but I recommend the previous review for a much more complete set of references. 
 
 Let us start by mentioning that in cosmology the use of the quantum chiral anomaly has been extensive, as all the baryogenesis models rely on it \cite{Sakharov:1967dj}. The CME and the chiral instabilities
 have also been used to explain the generation of primordial magnetic fields, with magnetic helicity (see, for example \cite{Joyce:1997uy,Boyarsky:2011uy}). There are also closed related
 works in cosmology and axions. Initially axions were proposed to solve the strong CP problem,  and nowadays they are  serious dark matter candidates 
\cite{Duffy:2009ig}. Axions and axion-like particles are predicted
 in many models as pseudo scalar particles that couple both to fermions and to photons. Most of the experiments to detect axions rely on how they couple to the electromagnetic fields.
 Several cosmological models of axions assume  that the axion field $a$ can be treated as a coherent classical field
 $ a(t) = \sqrt{2 \rho_{\rm DM}}{m_a} \sin{m_a t}$, where $\rho_{\rm DM}$ is the local dark matter density, and $m_a$ is the axion mass.
 Then, its time derivative acts as a chiral chemical potential for the fermions, and would imply that they produce a CME current.
 A new proposal to detect axions \cite{Hong:2022nss}, LACME (low temperature axion chiral magnetic effect), is based on this fact.
 
  Let me briefly mention how all these ideas are relevant for compact stars. There has not been the
 same amount of works on the impact of the chiral anomaly for compact stars than in cosmology, and most of the discussions are very qualitative, and simply do
  some estimates on the scales involved to assess how
 relevant chiral transport phenomena could be. Definitively, much more work is needed in this area of research.

It has been suggested  \cite{Ohnishi:2014uea} that the CME  could explain  the magnetic helicity of neutron stars, whose origin remains unknown.
In a proto neutron star one could create a big chiral misbalance in the population of electrons, which can be taken as quasi-massless as $\mu_e \sim 100$ MeV. The chiral misbalance
can be created in the neutronization process, as only left-handed leptons participate in this electroweak process. More particularly, when the neutron star is formed 
the  neutronization or electron capture
$ p + e_L\rightarrow n + \nu^e_L$, which is not yet counterbalanced with neutron decay, would lead to a high population of right-handed electrons.
Initial estimates of the chiral chemical potential
assumed these to be close to the QCD scale, creating a very large magnetic field, of the order of $B \sim 10^{18}$ G. However, it was later argued  \cite{Grabowska:2014efa} that 
the chirality flipping rate of Rutherford scattering of electrons off protons would  damp the chiral plasma instability. A more careful analysis \cite{Sigl:2015xva} however reveals that  electron capture rate  depends very much on the temperature, and that one can find for high $T$, but still for $\mu_e \gg T$,  that this mechanism is operating, creating magnetic fields of the order  $B \sim 10^{14}$ G.

Another set of ideas come in how the different field configurations of parallel electric and magnetic fields could create a chiral misbalance. In particular, these configurations occur in the
magnetospheres of supermassive black holes and of pulsars. In Ref.~\cite{Gorbar:2021tnw} estimates of the chiral densities generated in those cases are presented. While for supermassive
black holes the effect seems to be negligible, the authors of Ref.~\cite{Gorbar:2021tnw} suggest that it could be substantial for pulsars, estimating the chiral density created with the large time solution
of the chiral anomaly, with a chirality flipping rate due to electron-electron scattering. Then the effect should be detected by checking that the electromagnetic radiation in a given specific range of frequencies
should be circular polarized, as it propagates in a chiral medium.

It has also been considered that chiral effects might be relevant for the explanation of the so called pulsar kicks \cite{Hansen:1997zw}. Neutron stars typically have a velocity greater
than their progenitors, and many different sort of mechanism have been considered for the explanation of those kicks. The existence of the CME and CVE 
could lead to a possible explanation of the  kick \cite{Charbonneau:2009ax,Kaminski:2014jda}.

Chiral effects should be relevant to study the dynamics of neutrinos in core-collapse supernova. Most of the energy released in these supernova explosions is
in the form of neutrinos. The chirality of neutrinos should play a relevant role, and recently a chiral radiation transport theory has been put forward in
 Ref.~\cite{Yamamoto:2020zrs,Yamamoto:2021hjs,Matsumoto:2022lyb}.

\section{Helicity conservation law}
\vskip 0.5 cm

Studies of the relevance of chiral transport phenomena in compact stars are in their infancy, and typically focus on order of magnitude estimates.
The whole set of hydrodynamical equations, and not only the chiral anomaly, should be taken into consideration to asses its impact on the physics of
compact stars.

To this regard, an interesting claim has been made in the literature associated to classical hydrodynamics. There is an equation analogous to the chiral 
anomaly equation, valid for classical barotropic fluids \cite{Abanov:2021hio,Wiegmann:2022syo,Abanov:2022zwm}, which expresses the conservation law
of a combination of all the helicities that can be defined in the fluid. 
 In collaboration with Juan Torres-Rincon, we have
presented our own derivation in Ref.~\cite{Manuel:2022tck}, which extends the original derivation to the isentropic and finite temperature cases. It is easier to show it for relativistic fluids.
It only requires to derive hydrodynamical equations for the vectors $\omega^\mu$ and $B^\mu$ from the Euler equation obeyed by $u^\mu$, and use
some thermodynamical relations. More particularly, for isentropic fluids one finds
\begin{equation}
\label{central-eq}
   \partial_\mu ( h^2 \omega^\mu + h e B^\mu)= -  \frac{e^2}{4} \epsilon_{\mu \nu \alpha \beta} F^{\mu \nu} F^{\alpha \beta} \ ,
 \end{equation}
where $h$ is the enthalpy density. Thus, for isentropic fluids, there is an equation very similar to the chiral anomaly equation.
 This is a classical conservation
law, though. 
Let us stress that zero component of the $\omega^\mu$ vector
gives the fluid helicity, which measures the linkage of the fluid lines, and the zero component of the $B^\mu$ vector gives the mixed helicity, which measures
the linkage of magnetic and fluid lines. 
This equation expresses that there is a combination of the fluid, magnetic and mixed helicities which is conserved for isentropic fluids. As these
three helicities measure linkage, the equation   describes a  conservation of some topological properties of the system. 
It would be interesting to see how different dissipative effects might affect this conservation law.

We have also formulated how this equation is modified in the presence of a chiral imbalance, which then reads
\begin{equation}
\partial_\mu (h^2  \omega^\mu+h e B^\mu)
= - \frac{e^2}{4} \epsilon_{\mu \nu \alpha \beta} F^{\mu \nu} F^{\alpha \beta}+ (2h \omega^\mu + e B^\mu) (T\partial_\mu \bar{s}+\mu_5 \partial_\mu x_5) \ , 
\label{eq:conswithmu5}
\end{equation}
where $ \bar s =s/n$ is the entropy per particle,  $x_5 = n_5/n$ is the chiral fraction. Even for isentropic fluids ($\partial_\mu \bar{s}=0$), a space-time dependent chiral misbalance modifies the previous helicity conservation law. However, in the presence of chiral misbalance, Eq.~(\ref{eq:conswithmu5}) has to be combined with the chiral anomaly equation, Eq.~(\ref{chiral-anomaly}).
The chiral anomaly equation in the presence of a chiral chemical potential also involves the fluid and mixed helicities, through the axial vortical and
chiral separation effects. Sometimes the chiral anomaly equation has been taken as a sort of helicity conservation law \cite{Yamamoto:2015gzz}.

Let us stress that in all estimates of the relevance of the chiral anomaly presented in the previous section assumed that the fluid is at rest. 
However, this is certainly not the case in the astrophysical settings of interest, where we may expect the presence of vorticity and helicities, so it
might be interesting to revise all those estimates taking into consideration all hydrodynamical equations. 
More particularly, it would be interesting to review the generation of magnetic fields in the proto-nuetron stars, or the magnetic field dynamics in the magnetospheres
of magnetars.
We hope to report on this issue in the near future.

\ack
\vskip 0.5 cm
I thank the organizers of this wonderful and unique workshop for the invitation to give this talk and  for financial support.
This work was supported by Ministerio de Ciencia, Investigaci\'on y Universidades (Spain) under the project PID2019-110165GB-I00 (MCI/AEI/FEDER, UE), 
Generalitat de Catalunya by the project 2017-SGR-929 (Catalonia). This work was also partly supported by the Spanish program Unidad de Excelencia
 Maria de Maeztu CEX2020-001058-M, financed by MCIN/AEI/10.13039/501100011033

\end{document}